\documentclass[conference]{IEEEtran}
\IEEEoverridecommandlockouts

\usepackage{cite}
\usepackage{amsmath,amssymb,amsfonts}
\usepackage{algorithmic}
\usepackage{graphicx}
\usepackage{textcomp}
\usepackage{xcolor}
\usepackage{float}
\usepackage{booktabs} 
\usepackage{multirow}
\usepackage{array} 
\usepackage{caption} 
\usepackage{float}  
\usepackage{tabularx}
\usepackage{pifont}
\usepackage{hyperref}
\usepackage{subcaption}
\usepackage{verbatim}

\def\BibTeX{{\rm B\kern-.05em{\sc i\kern-.025em b}\kern-.08em
    T\kern-.1667em\lower.7ex\hbox{E}\kern-.125emX}}
\begin{document}

\title{ RSD-15K: A Large-Scale User-Level Annotated Dataset for Suicide Risk Detection on Social Media\\
}

\author{
\IEEEauthorblockN{Shouwen Zheng\IEEEauthorrefmark{1}\thanks{*All authors contributed equally.}}
\IEEEauthorblockA{\textit{Department of Computing} \\
\textit{The Hong Kong Polytechnic University}\\
Hong Kong \\
21097982d@connect.polyu.hk}
\and
\IEEEauthorblockN{Yingzhi Tao}
\IEEEauthorblockA{\textit{School of Computer Science and Technology} \\
\textit{Anhui University}\\
Hefei, China \\
e02114112@stu.ahu.edu.cn}
\and
\IEEEauthorblockN{Taiqi Zhou}
\IEEEauthorblockA{\textit{Department of Computing} \\
\textit{The Hong Kong Polytechnic University}\\
Hong Kong \\
21106717d@connect.polyu.hk}
}

\maketitle
\begin{abstract}
In recent years, cognitive and mental health (CMH) disorders have increasingly become an important challenge for global public health, especially the suicide problem caused by multiple factors such as social competition, economic pressure and interpersonal relationships among young and middle-aged people. Social media, as an important platform for individuals to express emotions and seek help, provides the possibility for early detection and intervention of suicide risk.
This paper introduces a large-scale dataset containing 15,000 user-level posts. Compared with existing datasets, this dataset retains complete user posting time sequence information, supports modeling the dynamic evolution of suicide risk, and we have also conducted comprehensive and rigorous annotations on these datasets. In the benchmark experiment, we systematically evaluated the performance of traditional machine learning methods, deep learning models, and fine-tuned large language models. The experimental results show that our dataset can effectively support the automatic assessment task of suicide risk. Considering the sensitivity of mental health data, we also discussed the privacy protection and ethical use of the dataset. In addition, we also explored the potential applications of the dataset in mental health testing, clinical psychiatric auxiliary treatment, etc., and provided directional suggestions for future research work.
\end{abstract}

\begin{IEEEkeywords}
Machine learning, natural language processing, suicide detection, mental health, dataset annotation
\end{IEEEkeywords}

\section{Introduction}
 With the popularity of social media, more and more individuals tend to express their psychological distress and suicidal thoughts on online platforms. Especially in the post-epidemic era, the sharp increase in life pressure has led to a significant increase in related cases\cite{Hamdan02012025}, making suicide risk detection on social platforms an urgent problem to be solved. However, how to accurately identify and assess suicide risk from massive social media data still faces major challenges.

Although there have been a lot of research in the field of suicide risk detection, there are still significant limitations in existing work\cite{Abdulsalam2024}. First, the size of the published datasets is generally small, which makes it difficult to support the training and verification of large-scale machine learning models. Second, existing datasets often lack unified annotation standards and mostly use simple binary classification annotations, which cannot reflect the complexity of suicide risk. In addition, most studies ignore the temporal evolution characteristics of user behavior and find it difficult to capture the dynamic changes of suicide risk. Most importantly, the annotation process generally lacks a strict quality control mechanism, which directly affects the reliability of the data and the validity of the research results. 

To solve the above problems, this paper proposes the RSD-15K dataset. This dataset contains about 15k user-level posts, which is one of the largest datasets of its kind. We not only retain the complete posting time sequence information of users to support risk evolution analysis, but also adopt a four-level risk level annotation system to significantly improve the precision of risk assessment. In the data annotation process, we implemented multiple rounds of cross-validation under professional guidance to ensure the reliability and consistency of the annotation results. This dataset is publicly available upon request\footnote{\url{https://github.com/Suicide-DataSet/RSD_15K}}

In the benchmark test, we used multiple baseline models from traditional machine learning to deep learning for system evaluation. The experimental results show that all kinds of models can achieve reliable performance in the user-level risk identification task, which strongly confirms the quality and practical value of the dataset. Especially in the modeling of time series features, our dataset shows unique advantages.

\begin{itemize}
\item we established the largest user-level suicide risk assessment dataset at present, and designed detailed annotation specifications and strict quality control processes to ensure professional annotation of all data;
\item we provided a complete baseline experimental framework, and verified the reliability and practical value of the dataset by systematically evaluating the performance of multiple machine learning models;
\item we deeply explored the issues of data privacy protection and ethical use, analyzed the key considerations in actual deployment, and provided directional suggestions for future research.
\end{itemize}

\section{Dataset processing}
		\subsection{Data Collection}
			\subsubsection{Raw Data Collection}
			All the data was extracted from a popular social media platform Reddit(\href{https://reddit.com}{https://reddit.com}, of which the submitted content is organized into different 'subreddits'. The individual communities enable us to collect data in certain topic. Specifically, we focus on the posts from 'suicidewatch' subreddit to ensure that they're suicide concerned. \par

            \begin{figure}[htbp]
    \centering
    \includegraphics[width=0.45\textwidth]{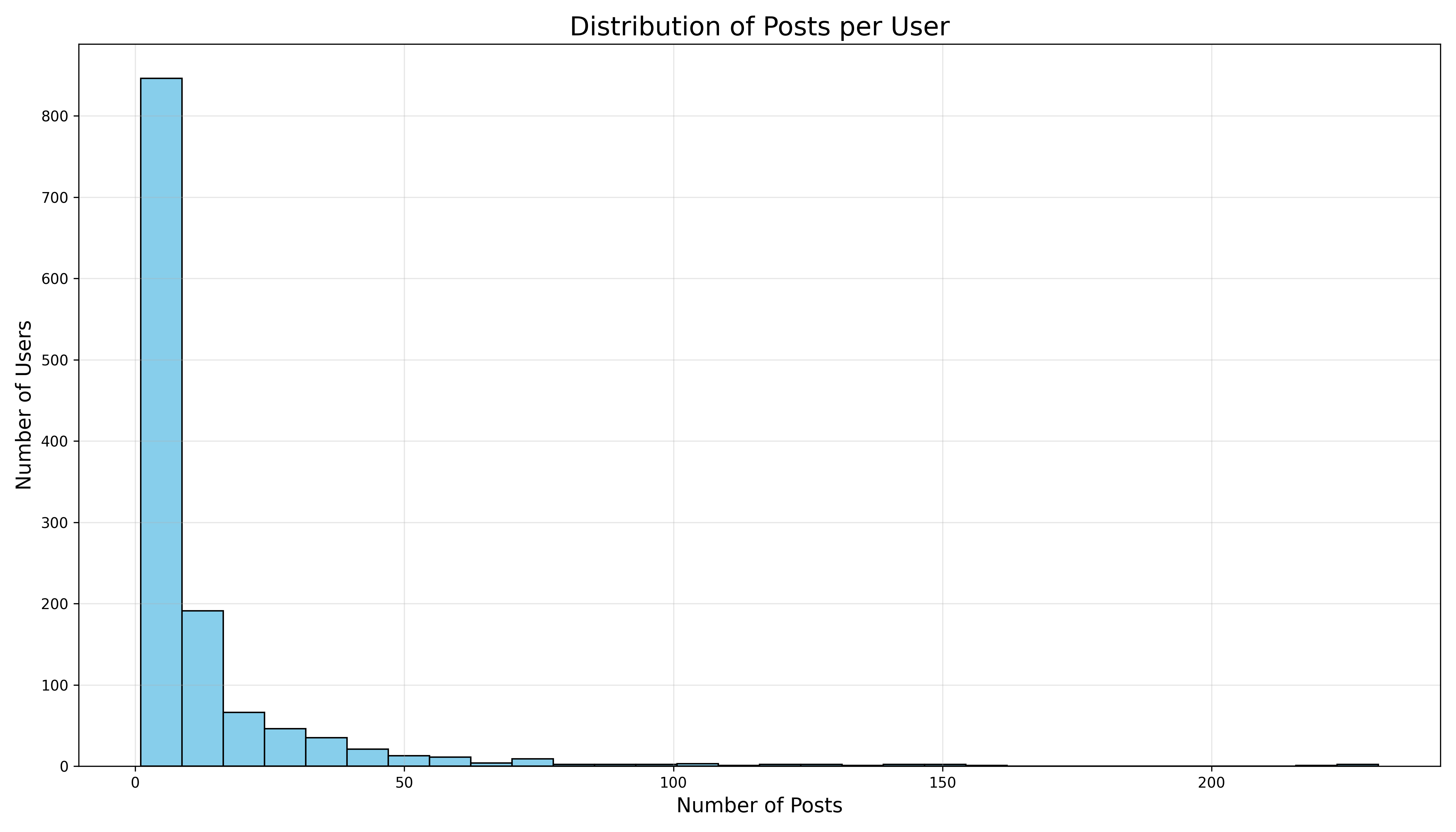}
    \caption{Distribution of Posts per User. It can be observed that the majority of users have fewer than 20 historical posts.}
    \label{fig:posts_per_user}
\end{figure}

			The data used in this study is sourced from the collection in \cite{li2022suicide} where 139,455 posts from 76,186 users published	under the “r/SuicideWatch” section from 01/2020 to 12/2021 have been crawled through the official Reddit API (as described in \cite{reddit_api}). However, in the origin work\cite{li2022suicide}, only 500 users' 3,998 posts were selected to be labeled.\par
							
			From the collected, but unannotated data, we selected 1,265 users' 14,613 posts for manual annotation, constructing out final dataset. This approach ensures the dataset is comprehensive, incorporating both relevant and non-relevant content, which provides a broad foundation for training machine learning models.

			\subsubsection{Pre-processing}
			For the pre-processing phase, several steps were taken to prepare the raw data for annotation. The data was first cleaned by removing non-relevant posts, such as those not related to the suicide risk theme. Duplicate posts were identified and removed to ensure data integrity. In addition, noise in the text, such as special characters, excessive punctuation, and irrelevant links, were filtered out to maintain focus on the relevant content. Tokenization and text normalization steps were carried out to standardize the text for machine learning applications. Further, the dataset was partitioned according to temporal constraints to facilitate time-series analysis, ensuring that posts were organized chronologically for proper risk evolution tracking.
			\subsubsection{Data Distribution}
\begin{figure}[htbp]
    \centering
    \includegraphics[width=0.5\textwidth]{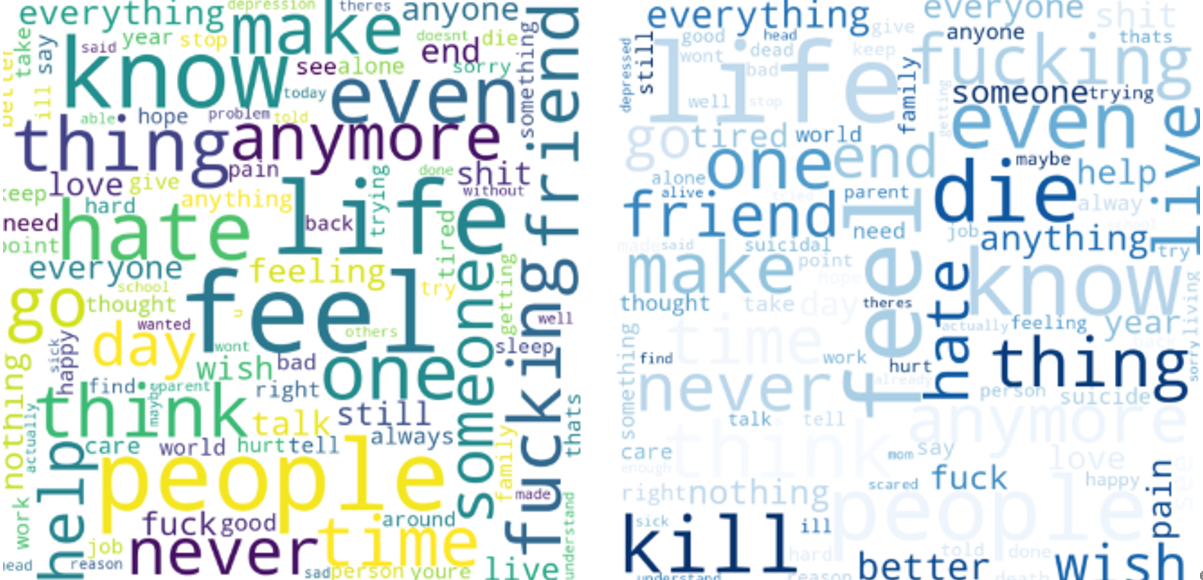}
    \caption{(a) Left: Indicator word cloud: n=4615 , (b) Right: Ideation: n=7133}
    \label{fig:wordclouds1}
\end{figure}

\begin{figure}[htbp]
    \centering
    \includegraphics[width=0.5\textwidth]{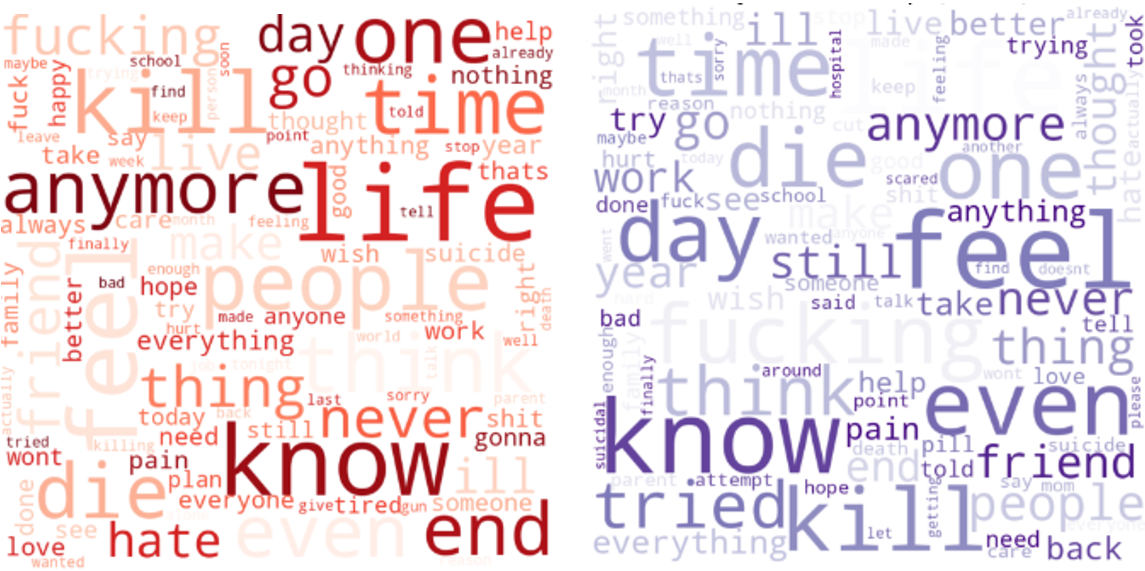}
    \caption{(a) Left: Behavior word cloud: n=2056 , (b) Right: Attempt word cloud: n=809}
    \label{fig:wordclouds2}
\end{figure}

\begin{figure}[htbp]
    \centering
    \includegraphics[width=0.45\textwidth]{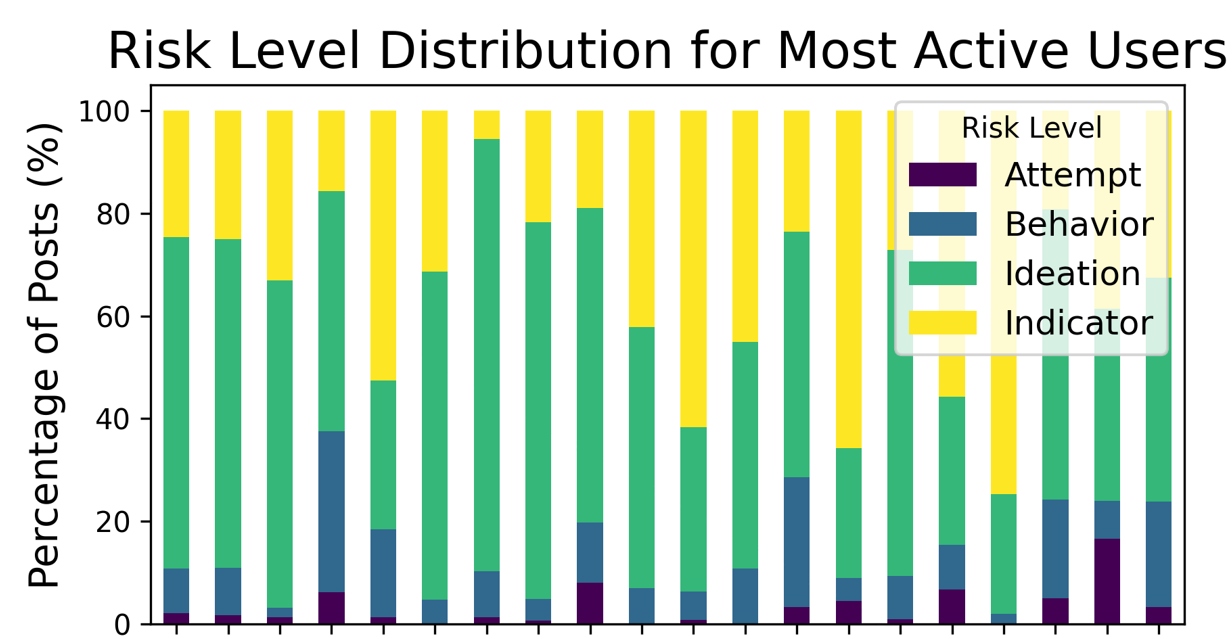}
    \caption{Risk Level Distribution for Most Active Users (Top 20). Note: User identifiers have been removed for privacy protection.}
    \label{fig:risk_level_distribution}
\end{figure}

			Collected data are all from social media and their users' information is anonymous, so the distribution details is unrelated to the users' personal identity information. The relationship between users and posts is firstly analyzed and shown in Figure \ref{fig:posts_per_user}. Overview in data is created using word-clouds grouped by labels. Indicator and Ideation are shown in Figure \ref{fig:wordclouds1}, while Behavior and Attempt shown in Figure\ref{fig:wordclouds2}. The data after annotation contains 14,613 valid data points, and detailed classes' distribution is shown in table \ref{tab:data_distribution}. To analyze typical users, the risk level distribution for the 20 most active users is shown in Figure \ref{fig:risk_level_distribution}. Since all the data is collected from real posts, a slight imbalance is both reasonable and authentic.
			\begin{table}[t]
				\centering
				\caption{Data Distribution}
				\begin{tabular}{|c|c|c|}
					\hline
					\textbf{Category} & \textbf{Count} & \textbf{Percentage} \\
					\hline
					Attempt & 809 & 5.54\% \\
					\hline
					Behavior & 2056 & 14.07\% \\
					\hline
					Ideation & 7133 & 48.81\% \\
					\hline
					Indicator & 4615 & 31.58\% \\
					\hline
				\end{tabular}
				\label{tab:data_distribution}
			\end{table}

		\subsection{Annotation}
			\subsubsection{Detailed Criteria}
			A detailed labeling criteria has been established in our work. Adapted from Columbia Suicide Severity Rating Scale (C-SSRS)\cite{posner2008columbia}, the dataset is categorized into four primary suicide risk labels: \textbf{Attempt}, \textbf{Behavior}, \textbf{Ideation}, and \textbf{Indicator}, a similar classifying categories with that in \cite{gaur2019knowledge} and \cite{li2022suicide}. To increase labeling consistency, we make more specific standard for annotation, including concrete detail.  \par 
			The \textbf{Attempt} label is assigned when the post mentions a previous suicide attempt, regardless of whether the author currently expresses suicidal thoughts. A suicidal attempt refers to a self-inflicted act intended to result in death, but which did not succeed. \par 
			The \textbf{Behavior} category includes preparatory acts or behaviors associated with self-harm or planning a suicide attempt. These behaviors go beyond mere verbalization and include any action such as purchasing tools, writing a suicide note, or preparing for death. Self-harm is also included in this category if it is not explicitly intended to end in death but involves self-inflicted harm. The \par 
			\textbf{Ideation} is classified when a post reflects suicidal thoughts or desires but lacks concrete actions. This includes both passive and active thoughts about suicide, such as wanting to die or expressing a desire to end one’s life. Unrealistic or hypothetical methods of suicide are also included, even if they cannot be directly acted upon. \par 
			The \textbf{Indicator} class is used for posts that do not indicate any suicidal risk. It includes posts referring to third parties’ suicidal behaviors or thoughts, as well as posts where the author denies any suicidal intent. This category is also used when the author expresses concern about another person's suicide risk.\par
			
			\subsubsection{Annotation Process}
			To increase the labeling efficiency, we make use of an open-source software Label Studio\cite{LabelStudio}, which is defined as a multi-type data labeling and annotation tool with standardized output format. The platform is able to conduct most labeling tasks and recognize most data formats. Our labeling task is conducted as a text classification task in community version. We deployed its Docker image on a server with Ubuntu20.04 OS from a commercial cloud service provider. The server has 8 cores with 16GB of RAM. The annotators access the platform via IP with port to carry out data labeling work.\par
			
			After confirming the labeling platform, three annotators are selected for labeling task. To ensure the quality in supervision, three supervisors are equipped. All the supervisors and annotators are knowledgeable in psychology domain and fluent in English. To ensure an optimal annotation option for voting when discrepancies occur, three people are selected, as three is the ideal number for balancing accuracy and efficiency.\par

			In this work, we mainly focus on actively preventing errors' existence instead of passively fixing or dropping them after occurrence. The reason is that the quantity of data is enormous and it's unrealistic to inspect every entry in the dataset. The reason is more convincing under the fact that except for the 30\% of the data that were jointly labeled for the calculation of Kappa\cite{fleiss1971measuring}, the remaining 70\% were labeled independently by each annotator(as discussed in \ref{sec:calculating_Kaapa}). Several \textbf{proactive measures} in avoiding generating errors are list in the following text and the \textbf{passive ongoing supervising measure} is also discussed in the last paragraph. \par
			
			Firstly, we train the annotators strictly before starting the main labeling task. After providing detailed labeling guidelines and instructions, 100 data samples were selected for expert annotation. The samples were utilized for verifying participants’ labeling accuracy before starting the formal task. If the accuracy from an annotator is below 95\%, the errors in the annotation are reviewed and corrected, followed by a re-annotation of the samples. This process continues until the accuracy reaches 95\%, after which the formal labeling process begins. Such training enables annotators fully understand the labeling rule and guarantees their accuracy.\par

			What's more, we innovatively set an \textbf{uncertainty reporting policy}. During formal labeling task, some cases are too ambiguous to be quickly classified, even relying on guesswork. In psychology domain, when humans are dealing with an uncertain task, they will be indecisive and are more easily to make bad decision. That is because when hesitation occurs, there will be confidence bias\cite{koriat2000feeling}\cite{boldt2017impact}, overthinking effect and confirmation bias\cite{nickerson1998confirmation}, making it harder to get the right judgment. To decrease labeling error rate in a low cost, we set an uncertainty reporting policy. Under this rule, when annotators meet with any uncertainty, they should leave the case without annotation and report it to supervisor. After one-day's work finished, such cases are gathered and used for joint decision-making to prevent errors. This policy guarantees labeling quality in a low cost.\par

			At the same time, to make sure that annotators can label with maximum efficiency and reduce the likelihood of errors, a daily annotation plan was established. Specifically, each annotator is assigned with 500 pieces of data a day. By allocating tasks effectively and setting clear work objectives, we can significantly enhance annotator productivity and minimize their error rate. Thus the quality and reliability of the dataset is guaranteed in another way.\par
			
			Despite mentioned preventive measures, a continuous monitoring measure is necessary.To supervise the quality during annotation process, we also set a daily inspection rule. Specifically, experts who set the standard randomly select 10\% of daily labeling results(after daily discussion) and examine them personally. Only when the accuracy of the reviewed data exceeds 85\%. Moreover, all reviews passed during the entire labeling phase, indirectly confirming the effectiveness of the proactive preventive measures mentioned above.
            
		\subsection{Quality Evaluation}
			\subsubsection{Annotation Consistency}
			\label{sec:calculating_Kaapa}
			For the consistency and accuracy of the annotations, we used Fleiss' Kappa\cite{fleiss1971measuring} as our objective verifying evaluation metric. To improve labeling efficiency, we chose about 30\% of the dataset, totaling 4,384 samples, for consistency testing and the remaining 70\%, totaling 10,229 data points, were labeled separately by three annotators. The resulting Kappa value of 0.7206 reflects a really good level of agreement among the annotators. We believe that the good annotation consistency is closely related to our extremely detailed and unambiguous annotation standard, similar backgrounds of annotators, uncertainty reporting policy, and the relaxed and stable annotation plan. For the collaboratively annotated 30\% data, we employed a three-person voting system for annotation. Any text where the annotators' annotations did not align with at least two identical responses were flagged for special review and re-annotation, ensuring that only high-quality and reliable data made it into the final dataset.		
		
			\subsubsection{Comparison with existing datasets}
			As shown in the table\ref{tab:dataset_comparison}, our dataset offers significant advantages over existing datasets in terms of granularity, annotation quality, accessibility, and data coverage. These key features make it a valuable resource for research in the field, providing a more robust foundation for model training and analysis compared to the datasets listed. Below, we highlight these advantages in greater detail.\par
			
			Firstly, in risk level part, our dataset is one of only two in this comparison that includes both post-level and user-level data. Furthermore, it has a larger sample size than the dataset in \cite{li2022suicide}. Additionally, our dataset includes timestamps for posts, which is valuable for time-series analyses and studies that require chronological context.\par
			Unlike the datasets in \cite{kaggle2021}, \cite{ji2018supervised}, \cite{cao2019latent}, and \cite{sinha2019suicidal}, which offer coarser categorizations, our dataset provides a more refined risk level classification. This higher granularity allows for better differentiation of risk levels, which is beneficial for training more accurate and effective models.\par
			Also, in contrast to the dataset in \cite{kaggle2021} and \cite{ji2018supervised}, which rely on community source for their labels, and the datasets in \cite{shing-etal-2018-expert} and \cite{zirikly-etal-2019-clpsych}, where most of the annotations are done by untrained crowdsourcing workers, our dataset is fully annotated by rigorously trained experts. This ensures higher annotation quality, reducing errors and enhancing the reliability of the labeled data for research purposes.\par
			What's more, our dataset is publicly available on the platform, adhering to the necessary regulations for access. This makes it more accessible for research than the dataset in \cite{sinha2019suicidal}, which, despite being fully expert-annotated, is not openly available. Researchers can easily apply to use our dataset, facilitating greater collaboration and progress in the field.\par

\begin{table*}[t]  
\centering  
\caption{Dataset Comparison}
\label{tab:dataset_comparison}  
\renewcommand{\arraystretch}{1.3} 
\scriptsize 

\begin{tabular}{>{\centering\arraybackslash}m{3.2cm} >{\centering\arraybackslash}m{1.2cm} 
                 >{\centering\arraybackslash}m{1.9cm} 
                 >{\centering\arraybackslash}m{1.7cm} 
                 >{\centering\arraybackslash}m{1.9cm} 
                 >{\centering\arraybackslash}m{2.0cm} 
                 >{\centering\arraybackslash}m{1.6cm} }
\hline
\multirow{2}{*}[-2.5ex]{\textbf{Dataset}} & \multirow{2}{*}[-2.5ex]{\textbf{Source}} & \multirow{2}{*}[-2.5ex]{\textbf{Size}} & \multicolumn{3}{c}{\textbf{Annotation}} & \multirow{2}{*}[-2.5ex]{\textbf{Availability}$^{(3)}$} \\ 
\cline{4-6}
 &  &  & \textbf{\begin{tabular}[c]{@{}c@{}}Risk Level$^{(1)}$\end{tabular}} & \textbf{\begin{tabular}[c]{@{}c@{}}Fine-grained SRL\end{tabular}} & \textbf{\begin{tabular}[c]{@{}c@{}}Fully Manual \\Annotation$^{(2)}$\end{tabular}} &  \\ 
\hline

Suicide and Depression Detection (Kaggle)\cite{kaggle2021} & Reddit & 
\begin{tabular}[c]{@{}c@{}}236,258 Posts\\ NO Users\end{tabular} & 
Post & \ding{55}$^{(4)}$ & \ding{55} & \ding{51}$^{(4)}$ \\ 
\hline

Suicidal Ideation Detection in Online User Content\cite{ji2018supervised} & 
Reddit, Twitter & 
\begin{tabular}[c]{@{}c@{}}7,098 Posts\\ 10,288 Tweets\\NO Users\end{tabular} & 
Post & \ding{55} & \ding{55} & \ding{55} \\ 
\hline

Latent Suicide Risk Detection on Microblog\cite{cao2019latent} & 
Tree Hole, Weibo &
\begin{tabular}[c]{@{}c@{}}744,031 Posts\\ 7,329 Users\end{tabular} & 
User & \ding{55} & \ding{51} & \ding{55} \\ 
\hline

Suicidal Ideation in Twitter\cite{sinha2019suicidal} & 
Twitter & 
\begin{tabular}[c]{@{}c@{}}34,306 Tweets\\ 32,558 Users\end{tabular} & 
Post & \ding{55} & \ding{51} & \ding{55} \\ 
\hline

Suicide Risk via Online Postings \cite{shing-etal-2018-expert} & 
Reddit & 
\begin{tabular}[c]{@{}c@{}}- Posts$^{(5)}$\\ 934 Users\end{tabular} & 
User & \begin{tabular}[c]{@{}c@{}}No / Low /\\Moderate / \\ Severe Risk\end{tabular} & \begin{tabular}[c]{@{}c@{}}\ding{51} \\ (mainly\\ crowdsourcing)\end{tabular} & \ding{51} \\ 
\hline

CLPsych2019\cite{zirikly-etal-2019-clpsych} & 
Reddit & 
\begin{tabular}[c]{@{}c@{}}- Posts\\ 621 Users\end{tabular} & 
User & \begin{tabular}[c]{@{}c@{}}No / Low /\\Moderate / \\ Severe Risk\end{tabular} & \begin{tabular}[c]{@{}c@{}}\ding{51} \\ (mainly\\ crowdsourcing)\end{tabular} & \ding{51} \\ 
\hline

Knowledge-aware Assessment of Suicide Risk \cite{gaur2021characterization} & 
Reddit & 
\begin{tabular}[c]{@{}c@{}}15,755 Posts\\500 Users\end{tabular} & 
User & Support / IN / ID / BR / AT$^{(6)}$ & \ding{51} & \ding{55} \\ 
\hline

Suicide risk level and trigger detection\cite{li2022suicide} & 
Reddit & 
\begin{tabular}[c]{@{}c@{}}3,998 Posts \\500 Users\end{tabular} & 
Post, User & IN / ID / BR / AT & \ding{51} & \ding{51} \\ 
\hline

Ours & 
Reddit & 
\begin{tabular}[c]{@{}c@{}}14,613 Posts \\1,265 Users\end{tabular} & 
Post, User & IN / ID / BR / AT & \ding{51} & \ding{51} \\ 
\hline
\end{tabular}

\vspace{0.2cm}
\begin{flushleft}
\footnotesize{
\textbf{Notes:} \\ 
(1) "User" in the "Risk Level" column indicates context-aware labels, while "Post" indicates that each post is individually labeled. \\ 
(2) "Fully manual annotation" means that each valid data unit has been manually reviewed, and no default labels have been applied due to the data source. \\ 
(3) "Availability" indicates whether the dataset can be accessed through public platforms based on certain regulations, without the need to contact the authors. \\ 
(4) \ding{55} denotes that there is no such setting in the datasets and \ding{51} indicates opposite. \\
(5) "- Posts" indicates that the dataset owner has not published or discussed the number of posts.\\
(6) IN: Indicator, ID: Ideation, BR: Behavior, AT: Attempt
}
\end{flushleft}
\end{table*}

\section{Benchmark Baselines and Evaluation}

In order to evaluate the performance of our proposed dataset on the task of "user-level suicide tendency prediction", this section selects five representative models as baselines for comparative analysis. These methods cover traditional machine learning, deep learning, and the currently popular Pre-trained Language Models (PLMs).\\
We divide suicide risk into four categories: indicator, ideation, behavior, and attempt. The suicide risk level of the user's latest post is used as the user's label. We give a time window to identify the user's suicide risk level. In this task, we mainly focus on the analysis of user sequential posts within a specific time window (including posting time and posting order, and the  stable version has 5 window elements). In terms of data partitioning, we randomly divide all users into training set (80\%), validation set (10\%), and test set (10\%) to ensure that the users from the training set and test set are entirely disjoint to prevent data leakage risks.\\

\subsection{Baseline Model}

\subsubsection{XGBoost}

We use XGBoost as a traditional machine learning baseline and build a multi-level feature engineering framework \cite{GHOSAL20231631}. It covers three dimensions: time, text, and sequence. In the time dimension, we analyze the temporal patterns of user posts, including posting interval statistics, time distribution, and behavior pattern characteristics; in the text dimension, we combine TF-IDF vectorization, text statistical features, and linguistic features; in the sequence dimension, we extract time series statistics, change trends, and historical cumulative features based on the historical post sliding window to capture the dynamic change pattern of user behavior.\\
Through feature importance analysis, we found that the time dimension features contribute most significantly to risk prediction, especially the change pattern of posting time intervals and the proportion of nighttime posts, indicating that the user's time behavior pattern is highly correlated with the risk level; and in the text features, the sudden change in content length also shows a strong predictive ability, which may reflect the change in the user's emotional state.\\

\subsubsection{BiLSTM}

We adopt a time-aware risk assessment model based on a bidirectional long short-term memory network (BiLSTM) to enhance the ability to express time-series features through a time encoding mechanism\cite{10.1007/978-981-99-6547-2_1}. The time encoding module uses three multi-dimensional encoding strategies: periodic time encoding, interval time encoding, and cumulative time features, to convert time information into dense vector representations to enhance the model's perception of time-series patterns.\\
The BiLSTM layer processes sequence data in both forward and backward directions to fully capture contextual information, allowing the model to consider both historical and future information. The model also introduces a multi-head attention mechanismThis mechanism integrates temporal features and text representation before BiLSTM, enabling the model to capture both internal text relations and temporal context information. This is different from the limitation of traditional BiLSTM that only processes sequence dependencies and cannot directly integrate external temporal features. This mechanism integrates temporal features and text representation before BiLSTM, enabling the model to capture both internal text relations and temporal context information. This is different from the limitation of traditional BiLSTM that only processes sequence dependencies and cannot directly integrate external temporal features.\\

\subsubsection{HiGRU}
We adopt a risk assessment model framework based on Hierarchical GRU\cite{jiao2019higruhierarchicalgatedrecurrent}, introduce a two-layer architecture, capture the semantic features of a single text through the bottom GRU, and model the temporal dependencies of user posting sequences through the top GRU. As the bottom structure, the text-level GRU converts the input text into a word embedding sequence, extracts features through the bidirectional GRU, and introduces residual connections and layer normalization mechanisms to improve training stability; while the user-level GRU constitutes the top structure, and handles long-term dependency issues through gating mechanisms and jump connections to retain and transmit historical information.\\
The model also integrates a time-aware attention mechanism, considering the time interval between posts, time periodicity features, and cumulative statistical features, and adaptively focuses on important historical information through the dynamic allocation of attention weights. The two-layer design enables the model to simultaneously focus on micro-semantic information and macro-behavioral patterns, effectively capturing the overall trend of individual post content and user posting sequences.\\

\subsubsection{RoBERTa}

We use the BERT framework as a fine-tuned large model baseline, specifically a time-aware model based on RoBERTa\cite{liu2019robertarobustlyoptimizedbert}, and introduce a temporal attention mechanism to enhance the model's modeling ability for user posting temporal features. In terms of temporal feature processing, we designed a multi-dimensional feature extraction scheme, including basic temporal features such as posting intervals and frequencies, periodic features of different scales such as days, weeks, and months, and temporal pattern features constructed through statistical analysis. These features are mapped to the same semantic space as the text representation through a specially designed temporal projection layer.\\
It is worth noting the design of the temporal attention mechanism, which adopts a multi-head attention structure, allowing the model to simultaneously focus on features of multiple time scales. Through residual connections and layer normalization, we ensure that temporal information can be effectively integrated with text representation while maintaining the integrity of the original semantic information; the calculation of attention weights takes into account the decay effect of temporal distance, enabling the model to more accurately capture temporal dependencies, while maintaining RoBERTa's powerful semantic understanding ability, and effectively integrating temporal information.\\

\subsubsection{DeBERTa}

DeBERTa has significant advantages in semantic understanding due to its enhanced debiased attention mechanism and relative position encoding\cite{he2021debertadecodingenhancedbertdisentangled}. We designed a multi-dimensional temporal feature encoding strategy, including the standardization of periodic features such as hours, weeks, dates, and months, the construction of time tag features such as night posting tags and weekend tags, and the mapping of time information to the same semantic space as the text representation through the feature projection layer, thus achieving effective fusion of temporal information.\\
In terms of text processing, we use the debiased attention mechanism to capture the subtle relationship between words, and the relative position encoding helps the model better understand the text structure. Through the temporal feature fusion strategy and multi-level feature extraction mechanism, the model can simultaneously grasp the semantic information of the text and the temporal pattern of user behavior, and accurately identify risk-related semantic clues.\\

\subsection{Experimental results and analysis}

\begin{table}[h]
\centering
\begin{tabularx}{0.5\textwidth}{lXXXXXX} 
\toprule
Model & Acc. (\%) & Mac-F1 (\%) & IN-F1 (\%) & ID-F1 (\%) & BR-F1 (\%) & AT-F1 (\%) \\ 
\midrule
XGBoost  & 42.5 & 25.3 & 58.2 & 37.6 & 39.0 & 31.2\\
BiLSTM   & 48.6 & 36.7 & 61.5 & 41.2 & 41.1 & 33.2\\
HiGRU    & 52.2 & 30.3 & 64.4 & 45.8 & 44.0 & 39.2\\
RoBERTa  & 71.0 & 65.0 & 72.0 & 73.7 & 72.0 & 71.0\\
DeBERTa  & 76.0 & 77.0 & 76.0 & 78.9 & 76.0 & 77.0\\
\bottomrule
\end{tabularx}
\caption{Performance comparison of baseline models on risk assessment task.}
\label{tab:baseline_performance}
\end{table}

Table III shows the performance comparison of baseline models on the risk assessment task. The results show a clear performance hierarchy between different model architectures. Transformer-based pre-trained language models significantly outperform traditional methods, with DeBERTa achieving the highest performance (76.0\% accuracy, 77.0\% macro F1), followed by RoBERTa (71.0\% accuracy, 65.0\% macro F1).

Sequence deep learning models show moderate performance, with HiGRU (52.2\% accuracy) and BiLSTM (48.6\% accuracy) both outperforming the traditional XGBoost method (42.5\% accuracy). This performance gradient is consistent with the increase in model complexity, demonstrating the value of advanced architectures in capturing complex patterns in risk assessment.

All models performed stably across multiple experimental runs, indicating high quality data annotation and reliable datasets. Performance across risk categories shows different sensitivities to model architecture, with transformer models showing balanced capabilities across all categories.

To verify the advantages of large-scale datasets in risk assessment tasks, we compared the performance of the same model architecture on datasets of different scales, as shown in Table 4.

\begin{table}[htbp]
\centering
\small
\setlength{\tabcolsep}{3pt}
\begin{tabular}{@{}l>{\raggedright\arraybackslash}p{1.5cm}ccccccc@{}}
\hline
\textbf{Data} & \textbf{Model} & \textbf{Opt.} & \textbf{In} & \textbf{ID} & \textbf{BR} & \textbf{AT} & \textbf{M-F1} & \textbf{Acc.} \\
\hline
500 & Large$^1$ & Full & 0.69 & 0.75 & 0.67 & 0.84 & 0.74 & 74\% \\
15K & Base$^1$ & No & 0.79 & 0.80 & 0.60 & 0.59 & 0.70 & 76\% \\
\hline
\multicolumn{9}{@{}p{\dimexpr\linewidth-2\tabcolsep}@{}}{\footnotesize{$^1$ DeBERTa Large (1.5B parameters) and Base (86M parameters) variants.}}\\
\end{tabular}
\caption{Comparison of F1 values of DeBERTa models across dataset sizes}
\label{tab:deberta_comparison}
\end{table}
In the small-scale dataset experiment, we used the DeBERTa-Large model to train on 500 annotated data, and adopted techniques such as hyperparameter optimization, data balance sampling, and model adjustment, and finally obtained an accuracy of 74\% and a macro-average F1 value of 0.74.In contrast, on the large-scale dataset we constructed containing 15,000 user-level annotated data, even using the DeBERTa-Base model with fewer parameters and without any hyperparameter adjustment or data balancing, it still achieved an accuracy of 76\% and a macro-average F1 value of 0.70.

The comparison chart proves that large-scale datasets can effectively compensate for the performance loss caused by the reduction of model parameters, this is consistent with previous research findings\cite{10.1371/journal.pone.0224365}; and without any optimization, the model on the large-scale dataset has surpassed the fully optimized small-scale dataset model\cite{10.1371/journal.pone.0224365}, indicating that the increase in data volume has a decisive influence on the improvement of model performance.

\section{Data privacy and ethical processing}
Given the sensitivity of the data we study, we took a number of steps to maintain privacy and ethical standards. In the dataset, all personal identifiers (such as usernames, specific post identifiers, and other metadata) were removed.[x1] After this anonymization process, there is no way to re-identify users from the data.

In addition, our dataset has been carefully constructed. All of our data only contains content that is voluntarily shared in public spaces (such as Reddit). Through these measures, we maintained the greatest possible respect for individual privacy. We ensured that this research is both scientifically valuable and ethically sound, laying the foundation for improved mental health interventions.

\section{conclusion}
The RSD-15K dataset is an important step forward in the direction of suicide risk detection on social media. With a large amount of user-level annotations, we provide a valuable resource for developing predictive models for mental health detection. The dataset contains unique temporal data and detailed risk level category combinations. Our annotation process remains rigorous to ensure the quality of the dataset. We also prioritize ethical and privacy issues to protect users' personal information. Future research can use this dataset to improve suicide detection systems and provide better support to those in need.

\bibliographystyle{IEEEtran}
\bibliography{references}

\end{document}